%% file: skeleton.tex
\documentclass[a4paper,11pt]{article}
\usepackage{pos}
\usepackage{bbold} 
\usepackage{todonotes} 
\usepackage{lineno} 

\title{Constraining Non-Standard Dark Matter-Nucleon Interactions with IceCube}
 \ShortTitle{Non-Standard Dark Matter Interactions with IceCube}

\author{The IceCube Collaboration \\{\normalsize \normalfont(a complete list of authors can be found at the end of the proceedings)}}

\emailAdd{lilly.peters@icecube.wisc.edu}

\abstract{After scattering off nuclei in the Sun, dark matter particles can be gravitationally captured by the Sun, accumulate in the Sun’s core and annihilate into Standard Model particles. Neutrinos originating from these annihilations can be detected by the IceCube Neutrino Observatory, located at the South Pole. Due to the non-observation of these neutrinos, constraints on the standard spin-dependent and spin-independent dark matter-nucleon scattering cross sections have been placed. Based on these constraints, we present upper limits on the coupling constants of the non-relativistic effective theory of dark matter-nucleon interactions, including velocity and momentum dependent interactions.\\

\vspace{4mm}
{\bfseries Corresponding authors:}
Lilly Peters$^{1,2*}$, Koun Choi$^{2}$, Mehr Un Nisa$^{3}$\\
{$^{1}$ \itshape III. Physikalisches Institut, RWTH Aachen University, D-52056 Aachen, Germany}\\
{$^{2}$ \itshape Dept. of Physics, Sungkyunkwan University, Korea}\\
{$^{3}$ \itshape Michigan State University}\\[4mm]
$^*$ Presenter
}

\FullConference{37$^{\rm{th}}$ International Cosmic Ray Conference (ICRC 2021)\\
		July 12th -- 23rd, 2021\\
		Online -- Berlin, Germany}

\begin{document}

\maketitle

\section{Introduction}
\input{1-introduction.tex}
\section{Theory}
\subsection{Effective Theory of Dark Matter-Nucleon Interactions}
\input{2-1-theory.tex} \label{sec:theory}
\subsection{Dark Matter Capture in the Sun}
\input{2-2-capture.tex} \label{sec:capture}
\section{Capture Rate Limits from IceCube Solar Dark Matter Analyses}\label{sec:icecube}
\input{3-icecube.tex}
\section{Results}\label{sec:results}
\input{4-results.tex}
\section{Conclusion and Outlook}
\input{5-conclusion.tex}

\bibliographystyle{ICRC}
\bibliography{references}


\clearpage
\section*{Full Author List: IceCube Collaboration}




\scriptsize
\noindent
R. Abbasi$^{17}$,
M. Ackermann$^{59}$,
J. Adams$^{18}$,
J. A. Aguilar$^{12}$,
M. Ahlers$^{22}$,
M. Ahrens$^{50}$,
C. Alispach$^{28}$,
A. A. Alves Jr.$^{31}$,
N. M. Amin$^{42}$,
R. An$^{14}$
% ...
\\

\noindent
% ...
$^{12}$ Universit{\'e} Libre de Bruxelles, Science Faculty CP230, B-1050 Brussels, Belgium \\
% ...
$^{14}$ Department of Physics and Laboratory for Particle Physics and Cosmology, Harvard University, Cambridge, MA 02138, USA \\
% ...
$^{17}$ Department of Physics, Loyola University Chicago, Chicago, IL 60660, USA \\
$^{18}$ Dept. of Physics and Astronomy, University of Canterbury, Private Bag 4800, Christchurch, New Zealand \\
% ...
$^{22}$ Niels Bohr Institute, University of Copenhagen, DK-2100 Copenhagen, Denmark \\
% ...
$^{28}$ D{\'e}partement de physique nucl{\'e}aire et corpusculaire, Universit{\'e} de Gen{\`e}ve, CH-1211 Gen{\`e}ve, Switzerland \\
% ...
$^{31}$ Karlsruhe Institute of Technology, Institute for Astroparticle Physics, D-76021 Karlsruhe, Germany  \\
% ...
$^{42}$ Bartol Research Institute and Dept. of Physics and Astronomy, University of Delaware, Newark, DE 19716, USA \\
% ...
$^{50}$ Oskar Klein Centre and Dept. of Physics, Stockholm University, SE-10691 Stockholm, Sweden \\
% ...
$^{59}$ DESY, D-15738 Zeuthen, Germany \\
% ...

\subsection*{Acknowledgements}

\noindent
USA {\textendash} U.S. National Science Foundation-Office of Polar Programs,
U.S. National Science Foundation-Physics Division,
U.S. National Science Foundation-EPSCoR,
Wisconsin Alumni Research Foundation,
Center for High Throughput Computing (CHTC) at the University of Wisconsin{\textendash}Madison,
Open Science Grid (OSG),
Extreme Science and Engineering Discovery Environment (XSEDE),
Frontera computing project at the Texas Advanced Computing Center,
U.S. Department of Energy-National Energy Research Scientific Computing Center,
Particle astrophysics research computing center at the University of Maryland,
Institute for Cyber-Enabled Research at Michigan State University,
and Astroparticle physics computational facility at Marquette University;
Belgium {\textendash} Funds for Scientific Research (FRS-FNRS and FWO),
FWO Odysseus and Big Science programmes,
and Belgian Federal Science Policy Office (Belspo);
Germany {\textendash} Bundesministerium f{\"u}r Bildung und Forschung (BMBF),
Deutsche Forschungsgemeinschaft (DFG),
Helmholtz Alliance for Astroparticle Physics (HAP),
Initiative and Networking Fund of the Helmholtz Association,
Deutsches Elektronen Synchrotron (DESY),
and High Performance Computing cluster of the RWTH Aachen;
Sweden {\textendash} Swedish Research Council,
Swedish Polar Research Secretariat,
Swedish National Infrastructure for Computing (SNIC),
and Knut and Alice Wallenberg Foundation;
Australia {\textendash} Australian Research Council;
Canada {\textendash} Natural Sciences and Engineering Research Council of Canada,
Calcul Qu{\'e}bec, Compute Ontario, Canada Foundation for Innovation, WestGrid, and Compute Canada;
Denmark {\textendash} Villum Fonden and Carlsberg Foundation;
New Zealand {\textendash} Marsden Fund;
Japan {\textendash} Japan Society for Promotion of Science (JSPS)
and Institute for Global Prominent Research (IGPR) of Chiba University;
Korea {\textendash} National Research Foundation of Korea (NRF);
Switzerland {\textendash} Swiss National Science Foundation (SNSF);
United Kingdom {\textendash} Department of Physics, University of Oxford.

\end{document}

%% file: 1-introduction.tex
Although there is compelling evidence for the existence of dark matter (DM), its nature remains unknown. To explain observations, a variety of theories provide candidate particles \cite{Bertone:2004pz}.
There are three complementary ways to search for DM - collider searches, direct and indirect detection. Direct detection aims to determine interactions of DM particles with nuclei, typically with the goal of setting constraints on spin-dependent (SD) and spin-independent (SI) interaction terms. However, there also are approaches to cover the full set of operators in a non-relativistic effective theory, including momentum- and velocity-dependent operators \cite{Fitzpatrick:2012ix}.
In recent years, there have been several efforts from direct detection experiments like CRESST-II \cite{Angloher:2018fcs}, Xenon100 \cite{Aprile:2017aas}, SuperCDMS \cite{Schneck:2015eqa} and DEAP-3600 \cite{Adhikari:2020gxw} to constrain the coupling constants of the theory.
With the IceCube Neutrino Observatory we are able to set limits via indirect detection based on DM capture in the Sun.

%% file: 2-1-theory.tex
We will briefly summarize the effective theory of DM-nucleon interactions following \cite{Fitzpatrick:2012ix}. The amplitude for elastic DM-nucleon scattering is restricted by several symmetries. In particular, Galilean invariance imposes that the scattering amplitude can only depend on combinations of the momentum transferred from the nucleon to the DM particle $\mathbf{q}$ and the relative incoming velocity $\mathbf{v} = \mathbf{v_{\chi}} - \mathbf{v}_{N}$ which is the incoming velocity of the DM particle $\chi$ in the rest frame of the nucleon $N$. In addition, the interactions must be hermitian, thus it is useful to work with hermitian quantities and to introduce $\mathbf{v}^\perp = \mathbf{v} + \mathbf{q}/2\mu_N$ with the reduced DM-nucleon mass $\mu_N$. All interaction operators can be written as a combination of $i\mathbf{q}$, $\mathbf{v^\perp}$, the DM spin $\mathbf{S}_{\chi}$ and the nucleon spin $\mathbf{S}_N$. They are listed in table \ref{tab:operators} imposing that they are at most linear in $\mathbf{S}_{\chi}$, $\mathbf{S}_N$ and $\mathbf{v^\perp}$. It is assumed that the mediating particle has spin 0 or spin 1 and is heavy compared to the momentum transfer.


The interaction Langrangian has the form
\begin{equation}\label{eq:lagrange}
    \mathcal{L}_{int} = \sum_{N=n,p} \sum_i c_i^N \mathcal{O}_i \chi^+\chi^- N^+N^-
\end{equation}

where $\chi^{\pm}$, $N^\pm$ are the fields involving only creation or annihilation fields and $c_i^p$ and $c_i^n$ denote the coupling constants for protons and neutrons and have mass dimension -2 \cite{Catena:2015uha}.
Instead of proton and neutron couplings we will present the results in terms of isoscalar $c^0 = c^p +c^n$ and isovector $c^1 = c^p -c^n$ couplings.

Forming a Lagrangian from the operators in table \ref{tab:operators} and performing a multipole expansion of the nuclear charges and currents lead to six nuclear response operators contributing to the transition probability, namely $M$, $\Sigma^{\prime}$, $\Sigma^{\prime\prime}$, $\Phi^{\prime\prime}$, $\tilde{\Phi^\prime}$ and $\Delta$.

Finally the differential cross-section for DM scattering off a nucleon of type N can be written as 
\begin{equation}\label{eq:crosssection}
    \begin{aligned}    
    \frac{d\sigma_N}{dE_R}(\omega^2, q^2) = \frac{2m_N}{\omega^2} \frac{1}{2J +1} & \sum_{\tau, \tau^\prime}
    \Big[ \sum_{k=M,\Sigma^{\prime}, \Sigma^{\prime\prime}}R_k^{\tau\tau^\prime}\Big(v_T^{\perp 2}, \frac{q^2}{m_N^2}\Big) \  W_k^{\tau\tau^\prime} (q^2) \\
    &+ \frac{q^2}{m_N^2} \sum_{k=\Phi^{\prime\prime},\Phi^{\prime\prime}M, \tilde{\Phi^\prime}, \Delta, \Delta\Sigma^\prime} R_k^{\tau\tau^\prime}\Big(v_T^{\perp 2}, \frac{q^2}{m_N^2}\Big) \  W_k^{\tau\tau^\prime} (q^2) \Big]
    \end{aligned}
\end{equation}
where the DM response functions $R_k^{\tau\tau^\prime}$ depend on the DM-nucleon interaction strength \cite{Catena:2015uha}.
The isotope-dependent nuclear response functions $W_k^{\tau\tau^\prime}$ have been calculated for the 16 most abundant elements in the sun through numerical shell model calculations in \cite{Catena:2015uha}.

\begin{table}
    \centering
    \begin{tabular}{ll}
        \hline
        $\mathcal{O}_1 = \mathbb{1}_{\chi N}$
        & $\mathcal{O}_{11} = i \mathbf{\hat{S}}_\chi \cdot \frac{\mathbf{\hat{q}}}{m_N} \mathbb{1_N}$\\
        $\mathcal{O}_3 = i \mathbf{\hat{S}}_N \cdot \Big(\frac{\mathbf{\hat{q}}}{m_N} \times \mathbf{\hat{v}}^\perp \Big) \mathbb{1_\chi}$
        & $\mathcal{O}_{12} = \mathbf{\hat{S}}_\chi \cdot \Big( \mathbf{\hat{S}}_N \times \mathbf{\hat{v}}^\perp \Big)$\\
        $\mathcal{O}_4 = \mathbf{\hat{S}}_\chi \cdot \mathbf{\hat{S}}_N$
        & $\mathcal{O}_{13} = i \Big( \mathbf{\hat{S}}_\chi \cdot \mathbf{\hat{v}}^\perp \Big) \Big( \mathbf{\hat{S}}_N \cdot \frac{\mathbf{\hat{q}}}{m_N} \Big)$\\
        $\mathcal{O}_5 = i \mathbf{\hat{S}}_\chi \cdot \Big(\frac{\mathbf{\hat{q}}}{m_N} \times \mathbf{\hat{v}}^\perp \Big) \mathbb{1_N}$
        & $\mathcal{O}_{14} = i \Big( \mathbf{\hat{S}}_\chi \cdot \frac{\mathbf{\hat{q}}}{m_N} \Big) \Big( \mathbf{\hat{S}}_N \cdot \mathbf{\hat{v}}^\perp \Big)$\\
        $\mathcal{O}_6 = \Big( \mathbf{\hat{S}}_\chi \cdot \frac{\mathbf{\hat{q}}}{m_N} \Big) \Big( \mathbf{\hat{S}}_N \cdot \frac{\mathbf{\hat{q}}}{m_N} \Big)$
        & $\mathcal{O}_{15} = - \Big( \mathbf{\hat{S}}_\chi \cdot \frac{\mathbf{\hat{q}}}{m_N} \Big) \Big[ \Big( \mathbf{\hat{S}}_N \times \mathbf{\hat{v}}^\perp \Big) \cdot \frac{\mathbf{\hat{q}}}{m_N} \Big]$\\
        $\mathcal{O}_7 = \mathbf{\hat{S}}_N \cdot \mathbf{\hat{v}}^\perp \mathbb{1_\chi}$
        & $\mathcal{O}_{17} = i \frac{\mathbf{\hat{q}}}{m_N} \cdot \mathcal{S} \cdot \mathbf{\hat{v}}^\perp \mathbb{1_N}$\\
        $\mathcal{O}_8 = \mathbf{\hat{S}}_\chi \cdot \mathbf{\hat{v}}^\perp \mathbb{1_N}$
        & $\mathcal{O}_{18} = i \frac{\mathbf{\hat{q}}}{m_N} \cdot \mathcal{S} \cdot \mathbf{\hat{S}}_N $\\
        $\mathcal{O}_9 = i \mathbf{\hat{S}}_\chi \cdot \Big( \mathbf{\hat{S}}_N \times \frac{\mathbf{\hat{q}}}{m_N}\Big)$
        & $\mathcal{O}_{19} = \frac{\mathbf{\hat{q}}}{m_N} \cdot \mathcal{S} \cdot \frac{\mathbf{\hat{q}}}{m_N} $\\
        $\mathcal{O}_{10} = i \mathbf{\hat{S}}_N \cdot \frac{\mathbf{\hat{q}}}{m_N} \mathbb{1_\chi}$& $\mathcal{O}_{20} = \Big( \mathbf{\hat{S}}_N \times \frac{\mathbf{\hat{q}}}{m_N} \Big) \cdot \mathcal{S} \cdot \frac{\mathbf{\hat{q}}}{m_N} $ \\
        \hline
    \end{tabular}
    \caption{\label{tab:operators}Non-relativistic quantum mechanical operators that are at most linear in $\mathbf{S}_{\chi}$, $\mathbf{S}_N$ and $\mathbf{v^\perp}$. $\mathcal{O}_2$ is quadratic in $\mathbf{v^\perp}$ and $\mathcal{O}_{16}$ a linear combination of $\mathcal{O}_{12}$ and $\mathcal{O}_{15}$, thus they are not considered here. $\mathcal{O}_{1}$ and $\mathcal{O}_{4}$ correspond to the standard SI and SD interactions. By introducing the nucleon mass $m_N$ all operators have the same mass dimension. Operators 17 - 20 can only arise for spin 1 DM and the symbol $\mathcal{S}$ denotes a symmetric combination of spin 1 polarisation vectors. In addition, $\mathcal{O}_{19}$ and $\mathcal{O}_{20}$ can only appear if the interaction is mediated by a vector mediator \cite{Catena:2019hzw}}
\end{table}

%% file: 2-2-capture.tex
DM particles of the Milky Way DM halo can scatter off nuclei in the Sun and be gravitationally captured if they scatter from a velocity $\omega$ to a velocity smaller than the local escape velocity $v_\odot^{\text{esc}}(r)$.

The capture rate for nuclei of type N is given by
\begin{equation}\label{eq:capturerate}
    C_{\text{cap}}^N = n_\chi \int_0^{R_\odot} dr \ 4 \pi r^2 n_N(r) \int_0^\infty du \ 4 \pi u^2 f_\odot (u) \frac{u^2 + v_\odot^{\text{esc}}(r)^2}{u} \int_{E_{min}}^{E_{max}} dE_R \ \frac{d\sigma_N}{dE_R} \ \theta(\Delta E)
\end{equation}
\begin{equation*}
    E_{\text{min}} = \frac{1}{2}M_\chi u^2 \qquad E_{\text{max}} = \frac{2\mu_N^2}{m_N}(u^2+v_\odot^{\text{esc}}(r)^2) \qquad \Delta E = E_{\text{max}} - E_{\text{min}}
\end{equation*}
where $u$ is the velocity of the DM particle at $r \rightarrow \infty$ in the solar frame such that $\omega = \sqrt{u^2 + v_\odot^{\text{esc}}(r)^2}$ \cite{Catena:2015uha}. We assume that the DM velocities follow a Maxwell-Boltzmann distribution in the galactic frame without a cut-off at the galactic escape velocity. We use $v_\odot=$ 220 km/s for the galactic orbital speed in the solar position, and $\sigma_v= 270$ km/s for the velocity dispersion.
To get the total capture rate $C_{\text{cap}}$, we sum equation (\ref{eq:capturerate}) for each of the 16 most abundant elements in the Sun. The elemental abundances $n_N$ are taken from the B16 GS98 model \cite{Vinyoles:2016djt}. Furthermore we assume the DM halo number density to be $n_\chi = 0.3 \ \text{GeV}/\text{cm}^3 /M_\chi$. For a discussion of the astrophysical uncertainties see section \ref{sec:syst_uncert}.

%% file: 3-icecube.tex
When DM particles are captured in the Sun, they can sink into the Sun's core and annihilate into Standard Model final states. Neglecting evaporation yields the following differential equation describing the number of DM particles in the Sun
\begin{equation}
 \frac{dN}{dt} = C_{\text{cap}} - C_{\text{ann}}N^2
\end{equation}
with the general solution $\Gamma_A = \frac{1}{2}C_{\text{cap}} \tanh^2{(\sqrt{C_{\text{cap}}C_{\text{ann}}}t)}$ \cite{Jungman:1995df}. Neutrinos originating from these annihilations can potentially be observed by neutrino telescopes like the IceCube Neutrino Observatory. IceCube is a cubic-kilometer sized neutrino detector installed in the ice at the geographic South Pole between depths of 1450 m and 2450 m, completed in 2010 \cite{Aartsen:2016nxy}. Neutrino reconstruction relies on the optical detection of Cherenkov radiation emitted by charged particles produced in the interactions of neutrinos in the surrounding ice or the nearby bedrock.

IceCube searches for an excess of neutrino events correlated with the direction of the Sun in order to detect signatures of Solar DM annihilation. The results published based on three years of data were able to constrain DM-nucelon scattering for three annihilation channels ($\tau^+ \tau^-$, $W^+ W^-$, $b\Bar{b}$) for DM masses up to 10 TeV \cite{Aartsen:2016zhm}. Recently, utilizing IceCube DeepCore data and new cuts, an updated analysis using seven years of data pushes IceCube sensitivity at the lower energy end of the DM mass range down to 5 GeV (publication under preparation), and also constrains DM annihilating directly to neutrinos.


%% file: 4-results.tex
\subsection{Capture rates}

To calculate the capture rate for each operator, the value of the  coupling constant was set to $c_i = 10^{-3} m_v^2$ with $m_v = 246.2$ GeV, while all other coupling constants were set to zero. The value is arbitrary as the capture rate is proportional to the coupling constant squared and the ratio of the two will be used as a conversionfactor in section \ref{subsec:limits}.
There are six isotopes that are responsible for the leading contribution for at least one interaction operator in a certain DM mass range, namely $^1$H, $^4$He, $^{14}$N, $^{16}$O, $^{27}$Al and $^{56}$Fe. The most important isotope for each operator is determined by a compromise between the elemental abundance, the properties of the nuclear response operators and kinematic factors that depend on the DM and the nucleon mass.



\subsection{Systematic uncertainties}\label{sec:syst_uncert}

The most prominent uncertainty in terms of astrophysical uncertainties is the local DM density that we assume to be 0.3 GeV/cm³, a widely-used value \cite{Nobile:2021qsn}. Recent measurements have best-fit values in the range 0.2 - 0.6 GeV/cm³ \cite{Nobile:2021qsn}. The capture rates and the limits in the next section can be adapted to a different density by scaling with the ratio of the two numbers.

\begin{figure}
\centering
\includegraphics[width=1.\linewidth]{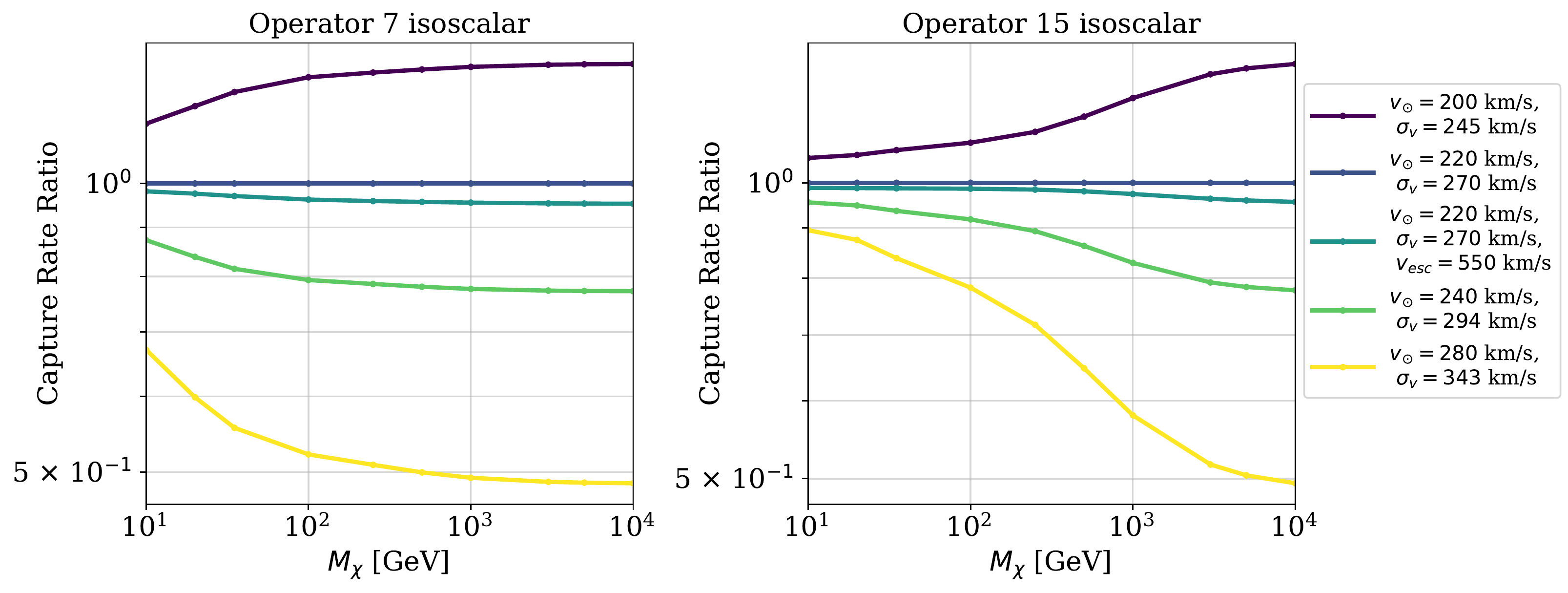}
\caption{\label{fig:syst_vdf}Capture rate ratio to the standard assumption as a function of the DM mass for different $f_\odot(u)$ with assumed $v_\odot$, $\sigma_v$ as specified in the legend.  The value of $\sigma_v$ is set to $\sqrt{3/2}v_\odot$. If no value for the galactic escape velocity is specified, the cut-off has been neglected. Exemplary, operator 7 and 15 are shown, dominated by hydrogen and iron, respectively.}
\end{figure}

The capture process is sensitive to deviations from the assumed Maxwellian velocity distribution. Figure \ref{fig:syst_vdf} shows the change in capture rate for a range of $v_\odot$ from 200 to 280 km/s, while $\sigma_v$ is set to $\sqrt{3/2} v_\odot$. Since Operator 7 is dominated by hydrogen, it is sensitive to changes in the velocity distribution earlier (at smaller DM masses) than operator 15, which is dominated by iron. 

Another uncertainty is the elemental composition of the sun. Especially the abundance of the respective dominant isotope is important. For a discussion of the uncertainties in the solar model see \cite{Vinyoles:2016djt}.

IceCube sensitivity to DM annihilation is also impacted by various detector systematics, which are included in the annihilation rate limits and propagate into the upper limits on the coupling constants.

\subsection{Limits on the Coupling Constants}\label{subsec:limits}

\begin{figure}
    \centering
    \includegraphics[width=1.\linewidth]{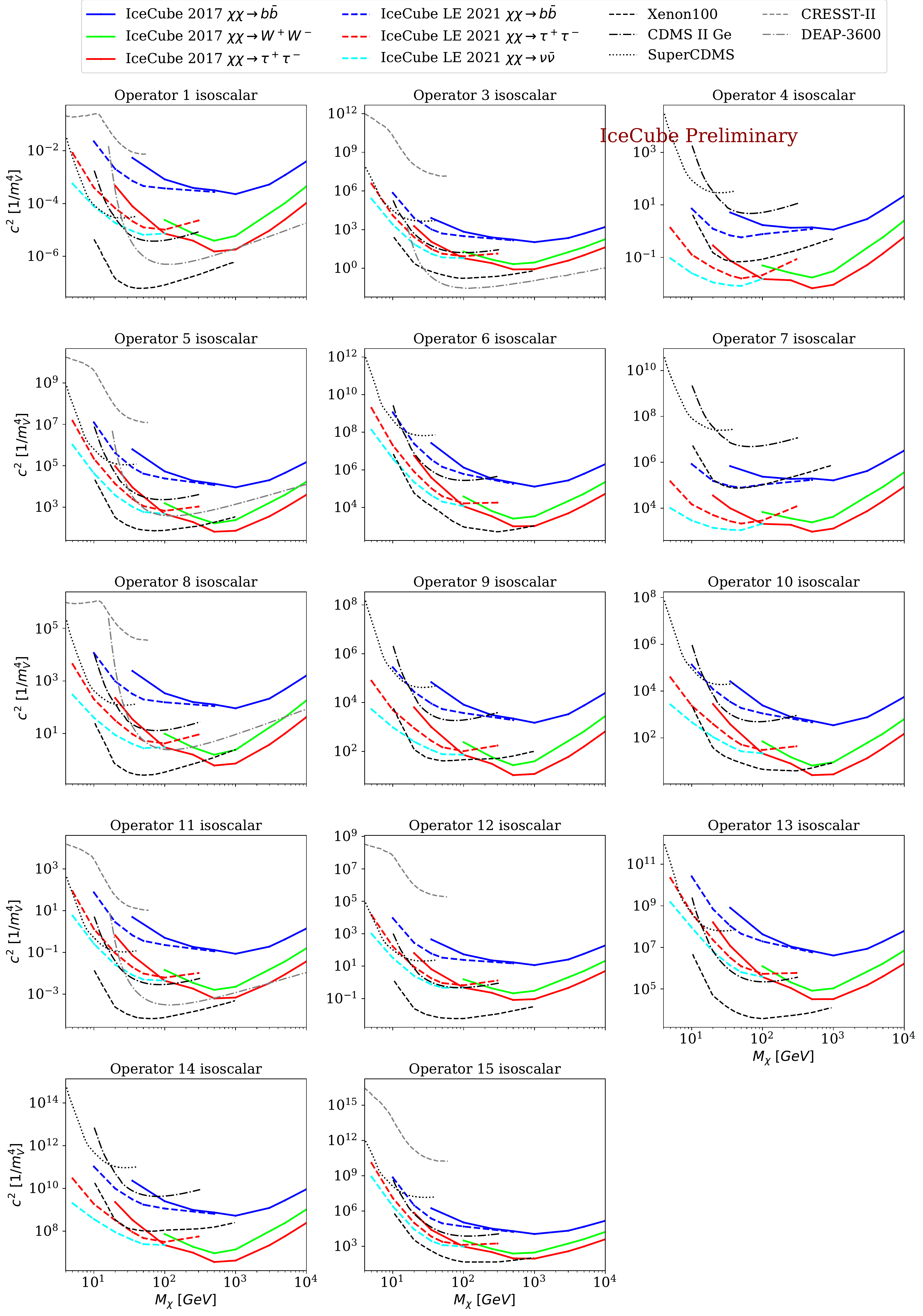}
    \caption{\label{fig:c2limits_isosca}Exclusion limits on the isoscalar coupling constants at 90 \% confidence level.}
\end{figure}

\begin{figure}
    \centering
    \includegraphics[width=1.\linewidth]{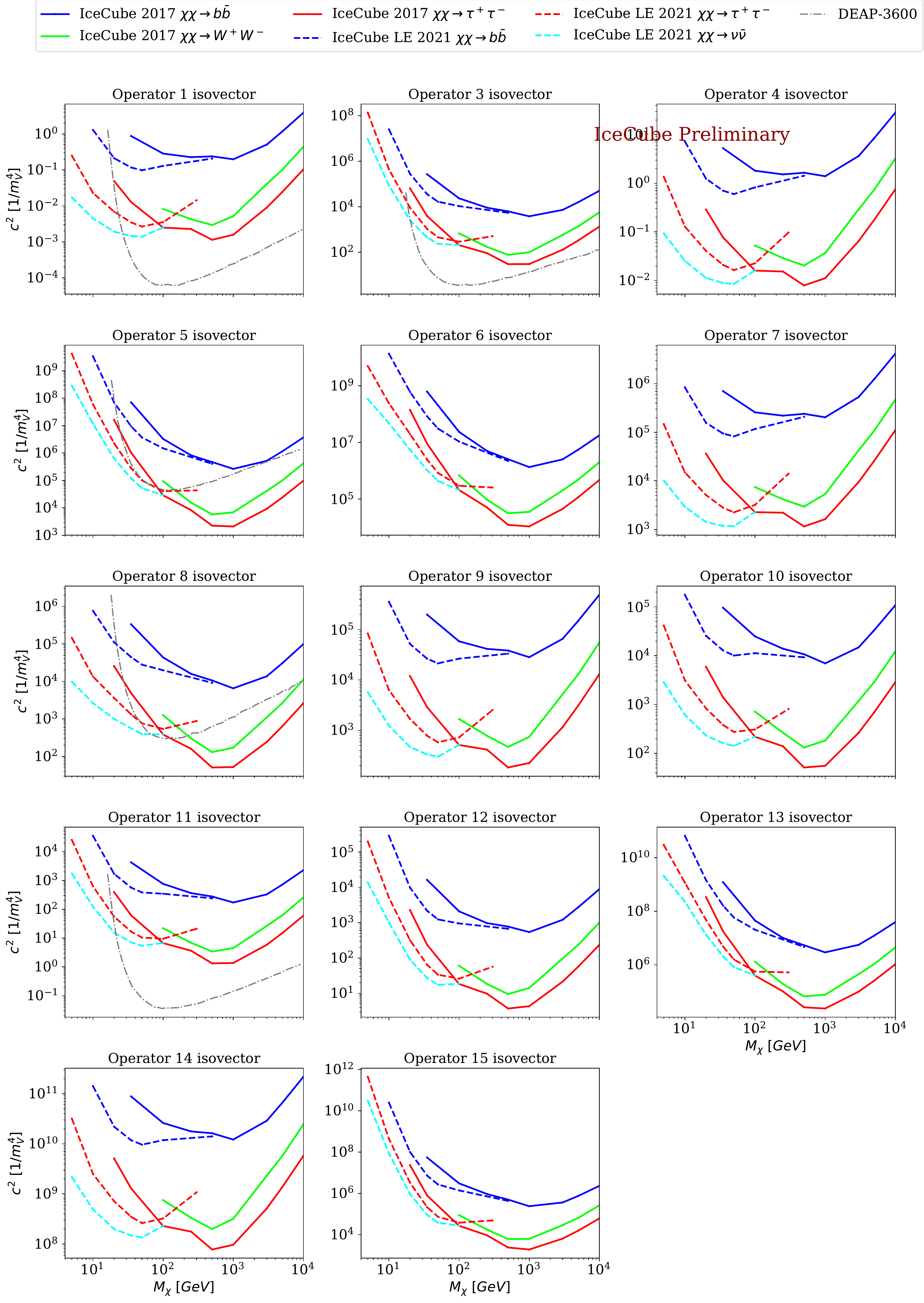}
    \caption{\label{fig:c2limits_isovec}Exclusion limits on the isovector coupling constants at 90 \% confidence level.}
\end{figure}

We can now use the computed capture rates to convert the capture rate limits from the IceCube analyses described in section \ref{sec:icecube} into limits on the coupling constants. The results are shown in figure \ref{fig:c2limits_isosca} and \ref{fig:c2limits_isovec} for isoscalar and isovector interactions. For simplicity we only present results for DM particles with spin $j_\chi = \frac{1}{2}$.
The limits are compared to the direct detection experiments CRESST-II \cite{Angloher:2018fcs}, Xenon100 \cite{Aprile:2017aas}, DEAP-3600 \cite{Adhikari:2020gxw}, CDMS II and SuperCDMS \cite{Schneck:2015eqa}. All exclusion limits are at 90 \% confidence level. CRESST-II is not competitive in the considered mass range but provides the strongest constraints below 5 GeV and is thus shown for completeness. A comparison to LUX \cite{Akerib:2021fxw} is not feasible since they present their results in terms of proton/neutron instead of isoscalar/isovector couplings. Because direct detection experiments are restricted to the isotopes in their detectors, they cannot always set constraints on all interaction types.
We observe that whether the strongest constraint comes from direct detection or IceCube is highly dependent on the operator. Notably, for Operator 4, 7 and 14 IceCube sets the most stringent limits for a large mass range, for the first two even with the soft channel analysis.

%% file: 5-conclusion.tex
We have reported upper limits on the isocalar and isovector coupling constants of the non-relativistic effective theory of DM-nucleon interactions for DM particles with spin 1/2. We want to stress that the theoretical framework presented in section \ref{sec:theory} is also valid for other spins and that the analysis can easily be extended.

The shown results are based on annihilation rate limits from two previous IceCube analyses, one from 2017 using three years of data \cite{Aartsen:2016zhm} and the other from 2021 focussing on the low energy regime (publication under preparation). Including new analyses like \cite{Lazar:2021icrc} could further improve the limits presented here.